\documentclass[10pt,conference]{IEEEtran}

\IEEEoverridecommandlockouts

\usepackage{algorithmic}
\usepackage{algorithm}

\usepackage[noadjust]{cite}
\usepackage{graphicx}

\usepackage{amsmath}   
\usepackage{amssymb}
\usepackage{amsfonts}

\usepackage{multirow}

\usepackage{dsfont}
\usepackage[latin1]{inputenc}
\usepackage{times}

\def\mb{\mathbf}
\def\opn{\operatorname*}
\def\bs{\boldsymbol}

\def\ds{\mathds}
\def\mc{\mathcal}
\def\ss#1{{\sf #1}}

\def\vec#1{\mb{#1}}
\def\rvec#1{\bs{\uppercase{#1}}} 
\def\rvecr#1{\bs{\lowercase{#1}}} 
\def\vecs#1{\bs{#1}}

\def\mat#1{\mb{\uppercase{#1}}}
\def\mats#1{\bs{#1}}

\def\dim{n}
\def\dimp{m}
\def\dimpp{p}

\def\R{\ds{R}}

\def\pdfvec#1{P_{\rvec{#1}}(\rvecr{#1})}

\def\EspOp{\ss{E}}

\newcommand{\Esp}[2][5]{%
  \ifcase#1
     \EspOp\{ #2 \}
     \or \EspOp \bigl\{ #2 \bigr\}
     \or \EspOp \Bigl\{ #2 \Bigr\}
     \or \EspOp \biggl\{ #2 \biggr\}
     \or \EspOp \Biggl\{ #2 \Biggr\}
  \else
     \EspOp \left\{ #2  \right\}
\fi}

\newcommand{\Earg}[3][5]{%
  \ifcase#1
     \EspOp_{#3} \{ #2 \}
     \or \EspOp_{#3} \bigl\{ #2 \bigr\}
     \or \EspOp_{#3} \Bigl\{ #2 \Bigr\}
     \or \EspOp_{#3} \biggl\{ #2 \biggr\}
     \or \EspOp_{#3} \Biggl\{ #2 \Biggr\}
  \else
     \EspOp_{#3} \left\{ #2  \right\}
\fi}

\newcommand{\CEsp}[3][5]{%
  \ifcase#1
     \EspOp\{ #2 \mid #3 \}
     \or \EspOp \bigl\{ #2 \bigm\vert #3 \bigr\}
     \or \EspOp \Bigl\{ #2 \Bigm\vert #3 \Bigr\}
     \or \EspOp \biggl\{ #2 \biggm\vert #3 \biggr\}
     \or \EspOp \Biggl\{ #2 \Biggm\vert #3 \Biggr\}
  \else
     \EspOp \left\{ #2  \,\middle\vert\, #3 \right\}
\fi}

\def\Jacob{{\ss D}}

\def\Tr{\ss{Tr}}
\def\det{\ss{det}}
\def\log{\ss{log}}

\def\vecop{\ss{vec}}

\def\pinv{\textrm{\footnotesize{+}}}

\def\d{\opn{d}\!}

\def\T{\ss{T}}

\def\I{I} 

\def\req#1{(\ref{#1})}

\def\eg{e.g.}

\newtheorem{prp}{Proposition}

\newtheorem{rem}{Remark}
\newtheorem{cor}{Corollary}
\newtheorem{lem}{Lemma}
\newtheorem{dfn}{Definition}

\def\Chan{\mat{H}}
\def\Prec{\mat{P}}
\def\UP{\mat{U}_\Prec}
\def\Singvals{\mats{\Sigma}}
\def\SP{\Singvals_\Prec}
\def\VP{\mat{V}_\Prec}
\def\QP{\mat{Q}_\Prec}
\def\NPrec{\bar{\Prec}}

\def\PT{\rho}
\def\Cov#1{\mat{R}_{\rvec{#1}}}
\def\Ssize{L}
\def\vecPrec{\vec{p}_1}
\def\EM{\mat{E}}
\def\MSE#1{\EM_{\rvec{#1}}}

\def\Sym#1{\mat{N}_{#1}}
\def\Dup#1{\mat{D}_{#1}}

\def\perfm{\mc{P}_0}

\begin{document}

\title{On optimal precoding in linear vector Gaussian channels with arbitrary input distribution}

\author{
\IEEEauthorblockN{Miquel Payaró}\thanks{This work was supported by
the RGC 618008 and the TEC2008-06327-C03-03/TEC research grants.}
\IEEEauthorblockA{Department of Radiocommunications \\
Centre Tecnològic de Telecomunicacions de Catalunya \\
Castelldefels, Barcelona, Spain \\
Email: {\tt miquel.payaro@cttc.cat}} \and \IEEEauthorblockN{Daniel
P.~Palomar}
\IEEEauthorblockA{Department of Electronic and Computer Engineering \\
Hong Kong University of Science and Technology \\  Clear Water Bay,
Kowloon, Hong Kong \\ Email: {\tt palomar@ust.hk}}}



\maketitle

\begin{abstract}
The design of the precoder the maximizes the mutual information in
linear vector Gaussian channels with an arbitrary input distribution
is studied. Precisely, the precoder optimal left singular vectors
and singular values are derived. The characterization of the right
singular vectors is left, in general, as an open problem whose
computational complexity is then studied in three cases: Gaussian
signaling, low SNR, and high SNR. For the Gaussian signaling case
and the low SNR regime, the dependence of the mutual information on
the right singular vectors vanishes, making the optimal precoder
design problem easy to solve. In the high SNR regime, however, the
dependence on the right singular vectors cannot be avoided and we
show the difficulty of computing the optimal precoder through an
NP-hardness analysis.
\end{abstract}

\section{Introduction}

In linear vector Gaussian channels with an average power constraint,
capacity is achieved by zero-mean Gaussian inputs, whose covariance
is aligned with the channel eigenmodes and where the power is
distributed among the covariance eigenva\-lu\-es according to the
waterfilling policy \cite{telatar:99, cover:91}. Des\-pite the
in\-for\-ma\-tion theoretic optimality of Gaussian inputs, they are
seldom used in practice due to their implementation complexity.
Rather, system designers often resort to simple discrete
constellations, such as BPSK or QAM.

In this context, the scalar relationship between mutual information
and minimum mean square error (MMSE) for linear vector Gaussian
channels put forth recently in \cite{guo:05}, and extended to the
vector case in \cite{palomar:06}, has become a fundamental tool in
transmitter design beyond the Gaussian signaling case.

In \cite{lozano:06}, the authors derived the optimum diagonal
precoder, or power allocation, in quasi-closed form, coining the
term mercury/waterfilling. Their results were found for the
particular case of a diagonal channel corrupted with AWGN and
imposing independence on the components of the input vector. The
mercury/waterfilling policy was later extended to non-diagonal
channels in \cite{f._perez-cruz_generalized_2007} through a
numerical algorithm.

The linear transmitter design (or linear precoding) problem was
recently studied in \cite{perezcruz:08,
rodrigues_multiple-input_2008} with a wider scope by considering
full (non-diagonal) precoder and channel matrices and arbitrary
inputs with possibly dependent components. In \cite{perezcruz:08,
rodrigues_multiple-input_2008} the authors gave necessary conditions
for the op\-ti\-mal precoder and optimal transmit covariance matrix
and proposed numerical iterative methods to compute a (in general
suboptimal) solution. Despite all these research efforts, a general
solution for this problem is still missing. In this work, we make a
step towards the characterization of its solution and give some
hints and ideas on why this problem is so challenging.

The contributions of the present paper are:
\begin{enumerate}
\item The expression for the optimal left singular vector matrix of the precoder that
maximizes a wide family of objective functions (including the mutual
information) is given.

\item We give a necessary and sufficient condition for the optimal singular values of the precoder that maximizes the
mutual information and propose an efficient method to numerically
compute it.

\item We show that the dependence of the mutual information on the
right singular vector matrix of the precoder is a key element in the
intractability of computing the precoder that maximizes the mutual
information.

\item We give an expression for the Jacobian of the mutual
information with respect to the transmitted signal covariance,
correcting the expression in \cite[Eq.~(24)]{palomar:06}.
\end{enumerate}


\emph{Formalism:} In this work we define a program according to
\begin{align}
\{ f_0^\star, x_1^\star, \ldots, x_{\dimp}^\star \}& = {\tt
Name}\left( a_1, \ldots, a_{\dimpp} \right) \nonumber \\ :=
\opn{max/min}_{x_1, \ldots, x_{\dimp}} & \:
f_0(x_1, \ldots, x_{\dimp}, a_1, \ldots, a_{\dimpp}) \\
\textrm{subject to} & \: f_i(x_1, \ldots, x_{\dimp}, a_1, \ldots,
a_{\dimpp}) \leq 0, \quad \forall i, \nonumber
\end{align}
where $( a_1, \ldots, a_{\dimpp} )$ are the parameters and $(x_1,
\ldots, x_{\dimp})$ are the optimization variables. Observe that the
first returned argument, $f_0^\star$, corresponds to the optimal
value of the objective function. We also make use of the Jacobian
operator $\Jacob$ applied to a matrix valued function $\mat{F}$ of a
matrix argument $\mat{X}$ defined as $\Jacob_{\mat{X}} \mat{F} =
(\partial \vecop \mat{F})/(\partial \vecop^\T \mat{X})$
\cite[Sec.~9.4]{magnus:88}, where $\vecop \mat{X}$ is the vector
obtained stacking the columns of $\mat{X}$. This notation requires
some modifications when either $\mat{F}$ or $\mat{X}$ are symmetric
matrices, see \cite{magnus:88} for details. In Section
\ref{sec:highSNR} we use some concepts of computational complexity
and program reductions. See \cite{papadimitriou_computational_1994,
goldreich_computational_2008} for reference.

\section{Signal model} \label{sec:model}

We consider a general discrete-time linear vector Gaussian channel,
whose output $\rvec{y} \in \R^{\dim}$ is represented by the
following signal model
%
%
\begin{gather} \label{eq:MIMO-prec}
\rvec{y} = \Chan \Prec \rvec{s} + \rvec{z},
\end{gather}
where $\rvec{s} \in \R^{\dimp}$ is the input vector distributed
according to $\pdfvec{s}$, the matrices $\Chan \in \R^{\dim\times
\dimpp}$ and $\Prec \in \R^{\dimpp \times \dimp}$ represent the
channel and precoder linear transformations, respectively, and
$\rvec{z} \in \R^{\dim}$ represents a zero-mean Gaussian noise with
identity covariance matrix $\Cov{z} = \mat{I}$\footnote{The
assumption $\Cov{z} = \mat{I}$ is made w.l.o.g., as, for the case
$\Cov{z} \neq \mat{I}$, we could always consider the whitened
received signal $\Cov{z}^{-1/2}\rvec{y}$.}.

For the sake of simplicity, we assume that $\Esp{\rvec{s}} =
\vecs{0}$ and $\Esp[1]{\rvec{s}\rvec{s}^\T} = \mat{I}$. 
The transmitted power $\PT$ is thus given by $\PT = \opn{Tr} \big(
\Prec \Prec^\T \big)$. We will also make use of the notation $\Prec
= \sqrt{\PT} \NPrec$, with $\opn{Tr} \big( \NPrec \NPrec^\T \big) =
1$ and also define $\Cov{\Chan} = \Chan^\T \Chan$. Moreover, we
define the SVD decomposition of the precoder as $\Prec = \UP \SP
\VP^\T$, the entries of $\SP$ as $\sigma_i = [\SP]_{ii}$, and also
the eigendecomposition of the channel covariance as $\Cov{\Chan} =
\mat{U}_{\Chan} \mats{\Lambda}_{\Chan}^2 \mat{U}_{\Chan}^\T$.
Finally, we define the MMSE matrix as $\EM_{\rvec{s}} =
\Esp[1]{(\rvec{s} - \CEsp{\rvec{s}}{\rvec{y}})(\rvec{s} -
\CEsp{\rvec{s}}{\rvec{y}})^\T}$.



\section{Problem definition and structure of the solution}
\label{sec:structure_solution}

In this paper we are interested in studying the properties of the
precoder $\Prec$ that maximizes the mutual information under an
average transmitted power constraint. However, in this section we
consider the more generic problem setup
\begin{align}
\{ \perfm^\star, \Prec_{\perfm}^\star \} = {\tt MaxPerformace} & \:
\big( \PT, \pdfvec{s}, \Cov{\Chan} \big) \nonumber \\ :=
\opn{max}_{\Prec} & \: \: \perfm \label{prg:MaxPerformance}
\\ \textrm{s.t.} & \: \: \opn{Tr} \big( \Prec \Prec^{\T} \big) =
\PT,
\nonumber
\end{align}
where $\perfm$ is a generic performance measure that depends on the
precoder $\Prec$ through the received vector $\rvec{y}$.

In the following lemma we characterize the dependence of $\perfm$ on
the precoder matrix $\Prec$.
\begin{lem} \label{eq:f_arb}
Consider a performance measure $\perfm$ of the system $\rvec{y} =
\Chan \Prec \rvec{s} + \rvec{z}$, such that $\perfm$ depends on the
distribution of the random observation $\rvec{y}$ conditioned on the
input $\rvec{s}$. It then follows that the dependence of $\perfm$ on
the precoder $\Prec$ is only through $\Prec^\T \Cov{\Chan} \Prec$
and we can thus write without loss of generality $\perfm =
\perfm\big( \Prec^\T \Cov{\Chan} \Prec \big)$.
\end{lem}
\begin{IEEEproof}
The proof follows quite easily by noting that $\Prec^\T \Chan^\T
\rvec{y}$ is a sufficient statistic of $\rvec{y}$, \cite[Section
2.10]{cover:91}. The sufficient statistic is thus $\Prec^\T \Chan^\T
\Chan \Prec \rvec{s} + \Prec^\T \Chan^\T \rvec{z}$. The first term
obviously depends on $\Prec$ only through $\Prec^\T \Cov{\Chan}
\Prec$. Since the second term $\Prec^\T \Chan^\T \rvec{z}$ is a
Gaussian random vector, its behavior is completely determined by its
mean (assumed zero) and its covariance matrix, given by $\Prec^\T
\Cov{\Chan} \Prec$.
\end{IEEEproof}

From all the possible choices for the performance measure function
$\perfm$, we are now going to focus our attention on the specific
class of \emph{reasonable} performance measures, which is defined
next.
\begin{dfn} \label{dfn:reasonable}
A performance measure $\perfm\big( \Prec^\T \Cov{\Chan} \Prec \big)$
is said to be reasonable if it fulfills that $\perfm\big( \alpha
\Prec^\T \Cov{\Chan} \Prec \big) > \perfm \big( \Prec^\T \Cov{\Chan}
\Prec \big)$, for any $\alpha > 1$ and $\Prec^\T \Cov{\Chan} \Prec
\neq \mats{0}$, which implies that $\perfm$ is a power efficient
performance measure.
\end{dfn}
\begin{rem}
The generic cost function\footnote{Observe that, while a performance
measure $\perfm$ is to be maximized, a cost function $f_0$ is
usually to be minimized.} $f_0$ considered in \cite{palomar:03} was
assumed to be a function of the elements of the vector
$\mathbf{diag}\big((\mat{I} + \Prec^\T \Cov{\Chan} \Prec)^{-1}\big)$
and increasing in each argument. Recalling that, for any $\alpha >
1$ and $\Prec^\T \Cov{\Chan} \Prec \neq \mats{0}$, we have
\begin{gather}
\big[ \mathbf{diag}\big((\mat{I} + \alpha \Prec^\T \Cov{\Chan}
\Prec)^{-1}\big) \big]_i < \big[ \mathbf{diag}\big((\mat{I} +
\Prec^\T \Cov{\Chan} \Prec)^{-1}\big) \big]_i.
\end{gather}
It is straightforward to see that the performance measure defined as
$\perfm \triangleq - f_0$ is a reasonable performance measure
according to Definition \ref{dfn:reasonable}.
\end{rem}


Based on a result in \cite{palomar:03} for the design of optimal
linear precoders, we characterize the left singular vectors of an
optimal precoder of \req{prg:MaxPerformance}.
\begin{prp} \label{prp:optU}
Consider the optimization problem in \req{prg:MaxPerformance}. It
then follows that, for any reasonable performance measure $\perfm$,
the left singular vectors of the optimal precoder $\Prec \in
\R^{\dimpp\times \dimp}$ can always be chosen to coincide with the
eigenvectors of the channel covariance $\Cov{\Chan}$ associated with
the $\min\{\dimpp, \dimp \}$ largest eigenvalues.
\end{prp}
\begin{IEEEproof}
For simplicity we consider the case $\dimp \geq \dimpp$. The case
$\dimp < \dimpp$ follows similarly. From the SVD of the precoder
$\Prec = \UP \SP \VP^\T$ and the eigen-decomposition of the matrix
\begin{gather}
\SP \mat{U}_\Prec^\T \Cov{\Chan} \mat{U}_\Prec \SP = \mat{Q}
\mats{\Delta} \mat{Q}^\T,
\end{gather}
with $\mats{\Delta}$ diagonal and $\mat{Q}$ orthonormal, it follows
that
\begin{gather}
\mat{Q}^\T \SP \mat{U}_\Prec^\T \Cov{\Chan} \mat{U}_\Prec \SP
\mat{Q}
\end{gather}
is a diagonal matrix. From \cite[Lemma 12]{palomar:03}, we can state
that there exists a matrix $\mat{M} = \mat{U}_\Chan
\mats{\Sigma}_{\mat{M}}$, with $\mats{\Sigma}_{\mat{M}}$ having
non-zero elements only in the main diagonal, such that $\mat{M}^\T
\Cov{\Chan} \mat{M} = \mats{\Delta}$ and that $\Tr \big(
\mat{M}\mat{M}^\T \big) \leq \Tr \big( \SP \big) = \Tr \big(
\Prec\Prec^\T \big)$. Now, we only need to check that
\begin{gather}
\Prec^\T \Cov{\Chan} \Prec 
= \mat{V}_\Prec \mat{Q}\mats{\Delta}\mat{Q}^\T \mat{V}_\Prec^\T =
\mat{V}_\Prec \mat{Q}\mat{M}^\T \Cov{\Chan} \mat{M}\mat{Q}^\T
\mat{V}_\Prec^\T . \nonumber
\end{gather}
Defining $\widetilde{\Prec} = \mat{M}\mat{Q}^\T \mat{V}_\Prec^\T =
\mat{U}_\Chan \mats{\Sigma}_{\mat{M}} \widetilde{\mat{V}}^\T$, with
$\widetilde{\mat{V}} = \mat{V}_\Prec \mat{Q}$, we have shown by
construction that for any given matrix $\Prec$ we can find another
matrix $\widetilde{\Prec}$ such that the objective function in
\req{prg:MaxPerformance} is the same,
\begin{gather}
\widetilde{\Prec}^\T \Cov{\Chan} \widetilde{\Prec} = \Prec^\T
\Cov{\Chan} \Prec \Rightarrow \perfm\big( \widetilde{\Prec}^\T
\Cov{\Chan} \widetilde{\Prec} \big) = \perfm\big( \Prec^\T
\Cov{\Chan} \Prec \big),
\end{gather}
which follows from Lemma \ref{eq:f_arb}, whereas the required
transmitted power is not larger, $\Tr \big( \widetilde{\Prec}
\widetilde{\Prec}^\T \big) = \Tr \big( \mat{M}\mat{M}^\T \big) \leq
\Tr \big( \Prec\Prec^\T \big)$. Since the performance measure
$\perfm$ is reasonable, the result follows directly.
\end{IEEEproof}

%

From the result in Proposition \ref{prp:optU}, it follows that, the
channel model in \req{eq:MIMO-prec} can be simplified, without loss
of optimality, to
\begin{gather} \label{eq:MIMO-prec-diag}
\rvec{y}' = \mats{\Lambda}_{\Chan} \SP \VP^\T \rvec{s} + \rvec{z},
\end{gather}
where now the only optimization variables are $\SP$ and $\VP$.

\section{Optimal singular values}

In this section we particularize the generic performance measure
considered in the previous section to the input-output mutual
information in \req{eq:MIMO-prec-diag}, i.e., $\perfm =  I(\rvec{s};
\rvec{y}')$. To compute the optimal $\SP^\star$ we define
\begin{align}
\{ I^\star, \SP^{2\star} \} = {\tt OptPowerAlloc} \: & \big( \PT,
\pdfvec{s}, \mats{\Lambda}_\Chan, \VP \big) \nonumber \\
:= \opn{max}_{\{\sigma_i^2\}} & \: I(\rvec{s}; \rvec{y}')
\label{prg:OptPowerAlloc}
\\ \textrm{s.t.} & \: \sum\nolimits_i \sigma_i^2 = \PT. \nonumber
\end{align}

%
Observe that the optimization is done with respect to the optimal
squared singular values. The optimal singular values are then
defined up to a sign, which does not affect the mutual information.
Consequently, we define $\sigma_i^\star = +\sqrt{\sigma_i^{2\star}}$
and $[\SP^\star]_{ii} = \sigma_i^\star$.

Let us now present an appealing property of $I(\rvec{s};
\rvec{y}')$.
\begin{lem}[\cite{payaro:09}] \label{lem:mi_concav}
Consider the model in \req{eq:MIMO-prec-diag} and fix $\VP$. Then it
follows that the mutual information $I(\rvec{s}; \rvec{y}')$ is a
concave function of the squared diagonal entries of 
$\SP$.
\end{lem}
With this result, we can now obtain a necessary and sufficient
condition for the squared entries of $\SP^\star$.
\begin{prp} \label{prp:Sstar}
The entries of the squared singular value matrix $\sigma_i^{2\star}
= [\SP^{2\star}]_{ii}$ of the solution to \req{prg:OptPowerAlloc}
satisfy
\begin{gather} \label{eq:KKT2}
\begin{split}
\sigma_i^{2\star} = 0 & \quad \Rightarrow \quad
[\mats{\Lambda}_\Chan^2]_{ii} \ss{mmse}_i(\SP^\star, \VP) < 2\eta \\
\sigma_i^{2\star} > 0 & \quad \Rightarrow \quad
[\mats{\Lambda}_\Chan^2]_{ii} \ss{mmse}_i(\SP^\star, \VP) = 2\eta,
\end{split}
\end{gather}
where $\eta$ is such that the power constraint is satisfied and
where we have used $\ss{mmse}_i(\SP^\star, \VP)$ to define the
$i$-th diagonal entry of the MMSE matrix $\MSE{\widehat{\rvec{s}}}$
corresponding to the model $\rvec{y}' = \mats{\Lambda}_{\Chan}
\SP^\star \widehat{\rvec{s}} + \rvec{z}$ with $\widehat{\rvec{s}} =
\VP^\T \rvec{s}$.
\end{prp}
\begin{IEEEproof}
The proof is based on obtaining the KKT conditions of the
optimization problem in \req{prg:OptPowerAlloc} together with
\begin{multline}
\frac{\d \I(\rvec{s} ; \mats{\Lambda}_{\Chan} \SP \widehat{\rvec{s}}
+ \rvec{z})}{\d (\sigma_i^2)} \\ = [\mats{\Lambda}_\Chan^2]_{ii}
\Esp{\left( \big[\widehat{\rvec{s}}\big]_i -
\CEsp[1]{\big[\widehat{\rvec{s}}\big]_i}{\mats{\Lambda}_{\Chan}
\SP^\star \widehat{\rvec{s}} + \rvec{z}}\right)^2},
\end{multline}
which follows from \cite[Cor.~2]{palomar:06}.
\end{IEEEproof}
\begin{rem}
The set of non-linear equations in \req{eq:KKT2} can be numerically
solved with, e.g., the Newton method because it has quadratic
convergence and the concavity property stated in Lemma
\ref{lem:mi_concav} guarantees the global optimality of the obtained
solution. The expression for the entries of the Jacobian vector of
$\ss{mmse}_i(\SP^\star, \VP)$ with respect to the squared entries of
$\SP$, which is needed at each iteration, is given by
\cite{payaro:09}
\begin{gather} \nonumber
\frac{\d \ss{mmse}_i(\SP^\star, \VP)}{\d (\sigma_j^2)} = -
[\mats{\Lambda}_\Chan^2]_{jj} \Esp{[\mats{\Phi}(\rvec{y}')]_{ij}^2},
\end{gather}
where $\mats{\Phi}(\rvecr{y}') =
\CEsp[2]{\widehat{\rvec{s}}\widehat{\rvec{s}}^\T}{\rvecr{y}'} -
\CEsp[2]{\widehat{\rvec{s}}}{\rvecr{y}'}\CEsp[2]{\widehat{\rvec{s}}^\T}{\rvecr{y}'}$.
\end{rem}

At this point, we have obtained the optimal left singular vectors
and the optimal singular values of the linear precoder that
maximizes the mutual information for a fixed $\VP$. Unfortunately,
the optimal solution for the right singular vectors $\VP$ seems to
be an extremely difficult problem. A simple suboptimal solution
consists in optimizing $\VP$ based on standard numerical methods
guaranteed to converge to a local optimum. See further
\cite{payaro:09a} for details on the practical algorithm to compute
the precoder. 

From the results presented in this section, it is apparent that the
difficulty of the problem in \req{prg:MaxPerformance} when
optimizing the mutual information lies in the computation of the
optimal right singular vectors matrix, $\VP^\star$. To support this
statement, in the following sections we deal with three cases: the
Gaussian signaling case, and the low and high SNR regimes. In the
Gaussian signaling case and low SNR regime, we recover the
well-known result that the mutual information depends only on the
squared precoder $\QP = \Prec \Prec^\T$ and is independent of the
right singular vectors matrix $\VP$, which further implies that, in
both cases, the optimal precoder can be easily computed. In Section
\ref{sec:highSNR} we will show that, for the high SNR regime, the
precoder design problem becomes computationally difficult through a
NP-hardness analysis.

\section{Situations where the mutual information is independent of $\VP$} \label{sec:lowSNR}

\subsubsection{Gaussian signaling case}

For the Gaussian signaling case, we recover the well known
expression for the mutual information \cite{cover:91}
\begin{gather}
I(\rvec{s};\rvec{y}) = \frac{1}{2}\log \det \big( \mat{I} + \QP
\Cov{\Chan} \big), \label{eq:I_Gaussian}
\end{gather}
from which it is clear that the only dependence of the mutual
information on the precoder is through $\QP = \UP \SP^2 \UP^\T$ and,
thus, it is independent of $\VP$. As we have pointed out in the
introduction and generalized in Proposition \ref{prp:optU}, the
optimal covariance $\QP$ is aligned with the channel eigenmodes
$\mat{U}_\Chan$. Also the power is distributed among the covariance
eigenva\-lu\-es $\SP^2$ according to the waterfilling policy
\cite{telatar:99}, which can be computed efficiently.

\subsubsection{Low SNR regime}

For the low SNR regime, a first-order expression of the mutual
information is \cite{palomar:06}
\begin{gather}
I(\rvec{s};\rvec{y}) = \frac{1}{2}\opn{Tr}\big( \QP \Cov{\Chan}
\big) + o\big( \Vert \QP \Vert \big). \label{eq:I_low_SNR}
\end{gather}
Just as in the previous case, from this expression it is clear that
the mutual information is insensitive to the right singular vector
matrix $\VP$. Moreover, the optimal matrix $\QP$ is easy to obtain
in closed form \cite{verdu:02}\footnote{The optimal signaling
strategy in the low SNR regime was studied in full generality in
\cite{verdu:02}. We recall that, in this work, we are assuming that
the signaling is fixed and the only remaining degree of freedom to
maximize the mutual information is the precoder matrix $\Prec$.}.

\begin{rem}
The expression in \req{eq:I_low_SNR} was derived in
\cite{palomar:06} through the expression of the Jacobian of the
mutual information with respect to $\QP$. Although the result in
\req{eq:I_low_SNR} is correct, the expression for the Jacobian
$\Jacob_{\QP} I(\rvec{s}; \rvec{y})$ given in
\cite[Eq.~(24)]{palomar:06} is only valid in the low SNR regime. The
correct expression for $\Jacob_{\QP} I(\rvec{s}; \rvec{y})$ valid
for all SNRs is \cite{payaro:08b}
\begin{multline} \label{eq:DqpI}
\Jacob_{\QP} \I(\rvec{s};\rvec{y}) = \frac{1}{2} \vecop^\T \big(
\Cov{\Chan} \Prec \mat{E}_{\rvec{s}} \Prec^{-1}  \big)  \Dup{\dim} \\
- \vecop^\T(\mat{E}_{\rvec{s}} \Prec^\T \Cov{\Chan} \UP \SP)
\mats{\Omega} \Sym{\dim} (\Prec^{-1} \otimes \Prec^\T)  \Dup{\dim},
\end{multline}
with
\begin{gather} \label{eq:DpVp}
\mats{\Omega} = \left( \begin{array}{c} \vec{v}_1^\T \otimes
\mat{V}_{\Prec}(\sigma_1^2\mat{I} - \mats{\Sigma}_{\Prec}^2)^\pinv
\mat{V}_{\Prec}^\T \\ \vec{v}_2^\T \otimes
\mat{V}_{\Prec}(\sigma_2^2\mat{I} - \mats{\Sigma}_{\Prec}^2)^\pinv \mat{V}_{\Prec}^\T \\
\vdots \\ \vec{v}_\dim^\T \otimes
\mat{V}_{\Prec}(\sigma_\dim^2\mat{I} -
\mats{\Sigma}_{\Prec}^2)^\pinv \mat{V}_{\Prec}^\T
\end{array}
\right),
\end{gather}
where $\vec{v}_i$ is the $i$-th column of matrix $\mat{V}_{\Prec}$,
$\Sym{\dim}$ and $\Dup{\dim}$ are the symmetrization and duplication
matrices defined in \cite[Secs. 3.7, 3.8]{magnus:88},
$\mat{A}^\pinv$ denotes the Moore-Penrose pseudo-inverse, and where
for the sake of clarity, we have assumed that $\Prec^{-1}$ exists
and that $\dim = \dimp = \dimpp$.
\end{rem}

\section{High SNR regime} \label{sec:highSNR}

In this section we consider that the signaling is discrete, i.e.,
the input can only take values from a finite set, $\rvec{s} \in
\mc{S} \triangleq \{ \vec{s}^{(i)} \}_{i=1}^\Ssize$. As discussed in
\cite{lozano:06, rodrigues_multiple-input_2008}, for discrete inputs
and high SNR, the maximization of the problem in
\req{prg:MaxPerformance} with the mutual information as performance
measure is asymptotically equivalent to the maximization of the
squared minimum distance, $d_{\min}$\footnote{Although we use the
symbol $d_{\min}$, it denotes squared distance.}, among the received
constellation points defined as $
d_{\min} = \min_{\vec{e} \in \mc{E}} \vec{e}^\T \Prec^\T \Cov{\Chan}
\Prec \vec{e}$, 
where $\mc{E}$ is the set containing all the possible differences
between the input points in $\mc{S}$.

Consequently, let us begin by considering the optimization problem
of finding the precoder that maximizes the minimum distance among
the received constellation points
\begin{align}
\{ d^\star, \Prec_d^\star \} = {\tt MaxMinDist} (\PT&, \mc{E},
\Chan) \nonumber \\ := \opn{max}_{\Prec} & \: \min_{\vec{e}
\in \mc{E}} \:\: \vec{e}^\T \Prec^\T \Cov{\Chan} \Prec \vec{e} \label{prg:MaxMinDist} \\
\textrm{s.t.} & \: \opn{Tr} \big( \Prec\Prec^\T \big) = \PT .
\nonumber
\end{align}

In the following, we give the proof that the program in
\req{prg:MaxMinDist} is NP-hard with respect to the dimension
$\dimp$ of the signaling vector, $\rvec{s}\in \R^\dimp$, for the
case where the set $\mc{E}$ is considered to be unstructured (i.e.,
not constrained to be a difference set). We are now preparing the
proof without this assumption in \cite{payaro:09a}. The proof is
based on a series of Cook reductions. We say that program ${\tt A}$
can be Cook reduced to program ${\tt B}$, ${\tt A}
\xrightarrow{\mbox{{\scriptsize \sc{Cook}}}} {\tt B}$, if program
${\tt A}$ can be computed with a polynomial time algorithm that
calls program ${\tt B}$ as a subroutine assuming that the call is
performed in one clock cycle. We have that, if ${\tt A}
\xrightarrow{\mbox{{\scriptsize \sc{Cook}}}} {\tt B}$ and ${\tt A}$
is NP-hard, then ${\tt B}$ is also in NP-hard,
\cite{goldreich_computational_2008}.

Before giving the actual proof we describe two more programs and
give some of their properties.

\subsection{Intermediate programs and their properties}

We first present the ${\tt MinNorm}$ program, which computes the
minimum norm vector that fulfills a set of constraints on its scalar
product with a given set of vectors $\{\vec{w}_{i} \}_{i =
1}^{\dimp}$
\begin{align}
\{ t^\star, \vec{z}^\star \} = {\tt MinNorm} \: \big(
\left\{\vec{w}_{i} \right\}_{i = 1}^{\dimp} & \big) \nonumber \\
:= \opn{min}_{\vec{z} \in \R^{\dimp}} & \: \Vert \vec{z} \Vert^2 \label{prg:MinNorm} \\
\textrm{s.t.} & \: |\vec{w}_{i}^\T \vec{z}| \geq 1, \quad i = 1,
\ldots, \dimp . \nonumber
\end{align}
\begin{lem}[\cite{luo_approximation_2007}]
${\tt MinNorm}$ is NP-hard.
\end{lem}

The second problem is ${\tt MinPower}$ and it computes the precoder
that minimizes the transmitted power such that the minimum distance
is above a certain threshold:
\begin{align}
\{ \PT^\star, \Prec^\star \} = {\tt MinPower}\left( d, \mc{E}, \Chan
\right) \nonumber \\ := \opn{min}_{\Prec} & \: \opn{Tr}
\big( \Prec\Prec^\T \big) \label{prg:MinPower} \\
\textrm{s.t.} & \: \min_{\vec{e} \in \mc{E}} \: \vec{e}^\T \Prec^\T
\Cov{\Chan} \Prec \vec{e} \geq d. \nonumber
\end{align}

\begin{lem} \label{lem:dualMinPowerMaxMinDist}
Assume that $\{d_0^\star, \Prec_0^\star\}$ is the output to the
program ${\tt MaxMinDist}(\PT_0, \mc{E}, \Chan)$. It then follows
that the output to ${\tt MinPower}\left( d_0^\star, \mc{E}, \Chan
\right)$ is given by $\{\PT_0, \Prec_0^\star\}$.

Similarly, assume that $\{\PT_0^\star, \Prec_0^\star\}$ is the
output to the program ${\tt MinPower}\left( d_0, \mc{E}, \Chan
\right)$. It then follows that the output to ${\tt
MaxMinDist}(\PT_0^\star, \mc{E}, \Chan)$ is given by $\{d_0,
\Prec_0^\star\}$.
\end{lem}
\begin{IEEEproof}
See \cite{boyd:04}.
\end{IEEEproof}

\begin{lem} \label{lem:ScaleMaxMinDist}
Assume that $\{d_0^\star, \Prec_0^\star\}$ is the output to the
program ${\tt MaxMinDist}(\PT_0, \mc{E}, \Chan)$. It then follows
that the output to ${\tt MaxMinDist}(\alpha \PT_0, \mc{E}, \Chan)$
with $\alpha > 0$ is given by $\{\alpha d_0^\star, \sqrt{\alpha}
\Prec_0^\star\}$.
\end{lem}
\begin{IEEEproof}
The proof follows easily, \eg, by considering the change of
optimization variable $\Prec = \sqrt{\alpha} \widetilde{\Prec}$ and
noting that the solution to the optimization problem remains
unchanged if the objective function is scaled by a constant
parameter.
\end{IEEEproof}

In the following we prove the following chain of reductions: ${\tt
MinNorm} \xrightarrow{\mbox{{\scriptsize \sc{Cook}}}} {\tt MinPower}
\xrightarrow{\mbox{{\scriptsize \sc{Cook}}}} {\tt MaxMinDist}$.

\subsection{Reduction of ${\tt MinNorm}$ to ${\tt MinPower}$}

In Algorithm \ref{alg:MinNormRed} we present our proposed Cook
reduction of ${\tt MinNorm}$ to ${\tt MinPower}$.
\begin{algorithm}[t]
\caption{Reduction of {\tt MinNorm} to {\tt MinPower}} \label{alg:MinNormRed}
\begin{algorithmic}[1]
\REQUIRE \hspace{.1cm} Set of weight vectors $\left\{ \vec{w}_{i}
\right\}_{i = 1}^{\dimp}$.

\ENSURE  Vector $\vec{z}^\star$ that achieves the minimum norm,
fulfill-\\ \hspace{.95cm} ing all the constraints $|\vec{w}_{i}^\T
\vec{z}^\star | \geq 1$. \\ \hspace{.95cm} Value of the minimum norm
$t^\star = \Vert \vec{z}^\star \Vert^2$.

\STATE \label{alg:H} Assign $\Chan = \left(
\begin{array}{cccc} 1 & 0 & \ldots & 0
\end{array} \right) \in \R^{1 \times \dimpp}$.

\STATE \label{alg:X} Assign $\mc{E} = \left\{ \vec{w}_{1}, \ldots,
\vec{w}_{\dimp} \right\}$.

\STATE \label{alg:CallMinPower} Call $\left\{ \PT^\star, \Prec^\star
\right\} = {\tt MinPower}(1, \mc{E}, \Chan)$.

\STATE $t^\star = \PT^\star$.

\STATE $\vec{z}^\star = ({\tt FirstRow}(\Prec^\star))^\T$.
\end{algorithmic}
\end{algorithm}

\begin{prp}
Algorithm \ref{alg:MinNormRed} is a polynomial time Cook reduction
of ${\tt MinNorm}$ to ${\tt MinPower}$.
\end{prp}
\begin{IEEEproof}
Under the assumption that ${\tt MinPower}$ can be solved in one
clock cycle, it follows that Algorithm \ref{alg:MinNormRed} runs in
polynomial time as well. It remains to check that the output of the
algorithm corresponds to the solution to ${\tt MinNorm}$.

Note that for the particular values assigned to the channel matrix
$\Chan$ and the set $\mc{E}$ in Steps \ref{alg:H} and \ref{alg:X} in
Algorithm \ref{alg:MinNormRed}, the program ${\tt MinPower}(1,
\mc{E}, \Chan)$ in \req{prg:MinPower} particularizes to
\begin{align} \label{eq:Part1MinPower}
\opn{min}_{\Prec} & \quad \opn{Tr} \big( \Prec\Prec^\T \big) \\
\label{eq:Part1MinPowerConst} \textrm{s.t.} & \quad \min_{i \in [1,
\dimp]} \: \vec{w}_{i}^\T \vecPrec \vecPrec^\T \vec{w}_{i} \geq 1,
\end{align}
where $\vecPrec$ is a column vector with the elements of the first
row of the precoder matrix $\Prec$. Observing that the constraint in
\req{eq:Part1MinPowerConst} only affects the elements of the first
row of matrix $\Prec$, it is clear that the optimal solution to
\req{eq:Part1MinPower} fulfills $[\Prec^\star]_{ij} = 0$, $\forall i
\neq 1$, as this assignment minimizes the transmitted power.
Recalling that $\vec{w}_{i}^\T \vecPrec \vecPrec^\T \vec{w}_{i} = |
\vec{w}_{i}^\T \vecPrec |^2$, it is now straightforward to see that
the first row of matrix $\Prec^\star$, which is the solution to the
problem in \req{eq:Part1MinPower}, is also the solution to ${\tt
MinNorm}$ in \req{prg:MinNorm}.%
\end{IEEEproof}
\begin{cor}
For the case where the set $\mc{E}$ is unconstrained, the program
${\tt MinPower}$ is NP-hard.
\end{cor}

\subsection{Reduction of ${\tt MinPower}$ to ${\tt MaxMinDist}$}

In Algorithm \ref{alg:MinPowerRed} we present our proposed Cook
reduction of ${\tt MinPower}$ to ${\tt MaxMinDist}$.

\begin{algorithm}[t] \caption{Reduction of {\tt MinPower} to {\tt MaxMinDist}} \label{alg:MinPowerRed}
\begin{algorithmic}[1]
\REQUIRE \hspace{.1cm} Desired squared minimum distance, $d$. \\
\hspace{0.95cm} Set of vectors $\mc{E}$. \\
\hspace{0.95cm} Channel matrix, $\Chan$.

\ENSURE Precoder $\Prec^\star$ that minimizes the transmitted power,
\\ \hspace{.95cm} fulfilling $ \min_{\vec{e} \in \mc{E}} \vec{e}^\T \Prec^{\star \T}
\Cov{\Chan} \Prec^\star \vec{e} \geq d$. \\
\hspace{.95cm} Transmitted power $\PT^\star = \opn{Tr} \big(
\Prec^{\star} \Prec^{\star \T} \big)$.

\STATE \label{alg:CallMaxMinDist} Call $\{ d_0^\star, \Prec_0^\star
\} = {\tt MaxMinDist}( 1, \mc{E}, \Chan)$.

\STATE Assign $\PT^\star = \frac{d}{d_0^\star}$.

\STATE Assign $\Prec^\star = \sqrt{\frac{d}{d_0^\star}}
\Prec_0^\star$.

\end{algorithmic}
\end{algorithm}

\begin{prp}
Algorithm \ref{alg:MinPowerRed} is a polynomial time Cook reduction
of ${\tt MinPower}$ to ${\tt MaxMinDist}$.
\end{prp}
\begin{IEEEproof}
Under the assumption that ${\tt MaxMinDist}$ can be solved in one
clock cycle, it follows that Algorithm \ref{alg:MinPowerRed} runs in
polynomial time as well. It remains to check that the output of the
algorithm corresponds to the solution to ${\tt MinPower}$.

Assume that the output to ${\tt MaxMinDist}(1, \mc{E}, \Chan)$ is
given by $\{d_0^\star, \Prec_0^\star\}$ as in Step
\ref{alg:CallMaxMinDist} in Algorithm \ref{alg:MinPowerRed}. Note
that, from the power constraint in \req{prg:MaxMinDist}, we have
that $\opn{Tr} \big( \Prec_0^\star \Prec_0^{\star \T} \big) = 1$.
From Lemma \ref{lem:ScaleMaxMinDist}, choosing $\alpha =
d/d_0^\star$, it follows that
\begin{gather}
\left\{d, \sqrt{d/d_0^\star} \Prec_0^\star \right\} = {\tt
MaxMinDist}\left( d/d_0^\star, \mc{E}, \Chan \right) .
\end{gather}
Now, applying Lemma \ref{lem:dualMinPowerMaxMinDist}, we have that
\begin{gather}
\left\{ d/d_0^\star, \sqrt{d/d_0^\star} \Prec_0^\star \right\} =
{\tt MinPower}( d, \mc{E}, \Chan),
\end{gather}
from which it immediately follows that $\PT^\star = d/d_0^\star$ and
$\Prec^\star = \sqrt{d/d_0^\star} \Prec_0^\star$, which completes
the proof. 
\end{IEEEproof}
\begin{cor}
For the case where the set $\mc{E}$ is unconstrained, the program
${\tt MaxMinDist}$ is NP-hard.
\end{cor}


Although the fact that the program ${\tt MaxMinDist}$ is NP-hard is
not a proof that the maximization of the mutual information is also
NP-hard, it gives a powerful hint on its expected computational
complexity in the high SNR regime where the minimum distance is the
key performance parameter.

From this expected complexity on the precoder design at high SNR and
the fact that, in Section \ref{sec:structure_solution}, we
characterized the optimal left singular vectors and the singular
values of the precoder that maximizes the mutual information as a
function \newpage \hspace{-.5cm} of the right singular vector matrix
$\VP$, it seems reasonable to place the computational complexity
burden of the optimal precoder design in the computation of
$\VP^\star$.

\section{Conclusion}

We have studied the problem of finding the precoder that maximizes
the mutual information for an arbitrary (but given) input
distribution. We have found a closed-form expression for the left
singular vectors of the optimal precoder and have given a sufficient
and necessary condition to compute the optimal singular values. We
have also recalled that, in the low SNR or Gaussian signaling
scenarios, the optimal precoder can be easily found as the mutual
information does not depend on the right singular vectors. Finally,
we have argued that in the high SNR regime, the computational
complexity of the calculation of the optimal right singular vectors
is expected to be hard.


%

%

\end{document}